\documentclass[12pt,preprint]{aastex}
\usepackage{natbib,psfig,amsmath}
\bibpunct{(}{)}{;}{a}{}{,}

\shorttitle{NIR Spectroscopy of McNeil's Nebula Object}
\shortauthors{Vacca et al.}

\begin{document}

\title{Near-Infrared Spectroscopy of McNeil's Nebula Object
\footnote{Based on observations made with the NASA Infrared Telescope Facility}}

\author{William D. Vacca}
\affil{SOFIA-USRA, NASA Ames Research Center, MS 144-2, Moffett Field, CA 94035}
\email{wvacca@mail.arc.nasa.gov}

\author{Michael C. Cushing\altaffilmark{2}}
\affil{SETI Institute, NASA Ames Research Center, MS 245-3, Moffett Field, CA 94035}
\email{mcushing@mail.arc.nasa.gov}
\altaffiltext{2}{Spitzer Fellow}

\and

\author{Theodore Simon}
\affil{Institute for Astronomy, University of Hawai`i, 2680 Woodlawn Drive, 
       Honolulu, HI 96822}
\email{simon@ifa.hawaii.edu}

\begin{abstract}

We present 0.8$-$5.2$\mu$m spectroscopy of the compact source at the base of a variable nebula (McNeil's Nebula Object) in the Lynds 1630 dark cloud that went into outburst in late 2003. The spectrum of this object reveals an extremely red continuum, CO bands at 2.3$-$2.5 $\mu$m in emission, a deep 3.0 $\mu$m ice absorption feature, and a solid state CO absorption feature at 4.7 $\mu$m. In addition, emission lines of H, Ca II, Mg I, and Na I are present. The Paschen lines exhibit P Cygni profiles, as do two lines of He I, although the emission features are very weak in the latter. The Brackett lines, however, are seen to be purely in emission. The P Cygni profiles clearly indicate that mass outflow is occurring in a wind with a velocity of $\sim 400$ km s$^{-1}$. The H line ratios do not yield consistent estimates of the reddening, nor do they agree with the extinction estimated from the ice feature ($A_V \sim 11$). We propose that these lines are optically thick and are produced in a dense, ionized wind.
The near-infrared spectrum does not appear similar to any known FUor or EXor object. However, all evidence suggests that McNeil's Nebula Object is a heavily-embedded low-mass Class I protostar, surrounded by a disk, whose brightening is due to a recent accretion event.

\end{abstract}
\keywords{stars: variables: other -- stars: pre-main sequence -- stars: formation }

\section{Introduction}

The object 2MASS J05461313$-$0006048 is located in the Lynds 1630 dark cloud of the Orion B molecular cloud/star formation complex. This very red object is associated with the IRAS source IRAS 05436$-$0007, the millimeter continuum source LMZ 12 (Lis, Menten, \& Zylka 1999), and the submillimeter continuum source OriB55smm (Mitchell et al.\ 2001; Johnstone et al.\ 2001). This object appears point-like in the deep $I-$band images of the region obtained by  Eisl\"offel \& Mundt (1997). Johnstone et al.\ (2001) estimated a size of $R_{\rm eff} \sim 6.8 \times 10^3$ AU and a mass of $0.68 M_\odot$ for the submillimeter source.  Based on the IRAS and millimeter data, Lis et al.\ (1999) estimated a bolometric luminosity for LMZ 12 of $2.7 L_\odot$ and identified it as a young, heavily embedded Class 0 source. This interpretation seemed to agree with the very red colors derived from the 2MASS colors, the fact that the source is invisible on the POSS plates, and the identification of the object by Eisl\"offel \& Mundt (1997) as the source driving the HH23 outflow. However, Lis et al.\ (1999) also noted that this classification was not completely consistent with the characteristics of other Class 0 sources, as LMZ 12 appeared to be very compact and lacked any sign of a molecular outflow.

This object recently experienced a dramatic brightening, 
associated with the sudden appearance of a conical-shaped nebula in optical images of the region (McNeil 2004). As the 2MASS source sits at the base of this reflection nebula, we shall henceforth refer to it as McNeil's Nebula Object (MNO). Reipurth \& Aspin (2004) have presented optical and near-infrared images of the source and its associated nebula, as well as optical and K-band spectra of MNO itself. Their spectra reveal strong $^{12}$CO emission, Br $\gamma$ in emission, and a P Cygni profile for the H$\alpha$ line. They suggest that the brightening of MNO may be related to a FUor- or EXor-like event. Brice\~no et al.\ (2004) have presented $I-$band images of MNO and its reflection nebula before and during the recent outburst. From optical spectra of MNO and a separate region in the nebula, obtained during the outburst, they deduce a spectral type of early B for MNO and a bolometric luminosity of $\sim 219 L_\odot$. Andrews, Rothberg, \& Simon (2004) have presented $10 \mu$m spectra, $^{12}$CO (2$-$1), 450 $\mu$m, and 850 $\mu$m observations of MNO in outburst. Integrating under the spectral energy distributions (SEDs) generated from all available pre- and post-outburst measurements, Andrews et al.\ (2004) estimate that the bolometric luminosity of MNO  increased by at least an order of magnitude, to $\sim 30-90 L_\odot$, during the outburst. Based on the shapes of the pre- and post-outburst SEDs, \'Abrah\'am et al.\ (2004) and Andrews et al.\ (2004) conclude that MNO is a Class I/Class II transition  object.

Here we present near-infrared (0.8$-$5.2$\mu$m) medium resolution spectroscopy of MNO obtained during the outburst, approximately one month after the observations of Reipurth \& Aspin (2004) and 1-2 days before the $10\mu$m and submillimeter observations of Andrews et al.\ (2004).  Based on these data, we suggest that MNO is a low-mass late-stage Class I protostar with mass outflow occurring in a dense, ionized wind. In this paper we assume a distance to MNO of 400 pc (Antony-Twarog 1982).

\section{Observations and Data Reduction}

MNO was observed at the NASA Infrared Telescope Facility on Mauna Kea  on 2004 March 09, 06h UT with SpeX, the facility near-infrared medium-resolution cross-dispersed spectrograph (Rayner et al.\ 2003). Ten individual exposures of MNO, each lasting 120 s, were obtained using the short wavelength cross-dispersed mode (SXD, covering 0.8$-$2.5 $\mu$m) of SpeX. These were followed by ten exposures, each lasting 30 s, with the long wavelength cross-dispersed mode (LXD2.3, covering 2.3$-$5.5$\mu$m). The observations were acquired in  ``pair-mode", in which the object is observed at two separate positions along the slit. We used the $0\farcs3$-wide slit, which yields nominal resolving powers of 2000 and 2500 for the SXD and  LXD2.3 modes, respectively. The slit was set to the parallactic angle for the SXD observations. Observations of an A0V star, used as a ``telluric standard", were obtained immediately preceding or following the observations of MNO. A set of internal flat fields and arc frames (for wavelength calibration) were also obtained for each mode. The airmass varied between 1.1 and 1.4 during the observations; however the airmass differences between the object and the standard were less than $0.06$. The seeing was estimated to be $\sim 1\arcsec$ at 2.2 $\mu$m and conditions were clear. 

The data were reduced using the Spextool package, which performs flat-fielding, non-linearity corrections, pair subtraction, aperture definition, optimal spectral extraction, and wavelength calibration of SpeX data semi-automatically (see Cushing et al.\ 2004).
Flux calibration of the spectra was carried out using the method and software described by Vacca et al.\ (2003);  the flux calibration is expected to be accurate to about $10$\%. After flux calibration, the individual SXD and LXD2.3 spectral orders were spliced together. 
The final spectrum spanning 0.8$-$5.2 $\mu$m is shown in Fig.\ 1. An enlargement of the 1$-$1.3 and 1.9$-$2.5 $\mu$m regions, after normalization of the continuum with a low order polynomial, is also shown. The signal-to-noise ratio (S/N) varies across the spectral range, but is of the order of 200 and 150 at the centers of the J and K bands, respectively. At the longest wavelengths the S/N of the MNO spectrum is limited by that of the A0V telluric standard spectrum (S/N $\sim 40$). 

\section{Results}

The 0.8$-$5.2 $\mu$m spectrum of MNO exhibits a very red continuum with a number of emission and absorption features, the most prominent of which are the broad ice absorption feature at 3.1 $\mu$m, the deep CO solid-state absorption feature at 4.7 $\mu$m, the CO ro-vibrational $\Delta\nu=+2$ transitions at 2.3$-$2.4$\mu$m in emission, and the Pa $\alpha$, Pa $\beta$, Br $\alpha$, and Br $\gamma$ emission lines. Also present are weaker emission features of Ca II at $0.85-0.86$ $\mu$m, Fe I at $1.16-1.20$ $\mu$m, Mg I at 1.50 $\mu$m, Na I at 2.21 $\mu$m, and Ca I at 2.26 $\mu$m. Although the Brackett lines appear to be purely in emission, the Pa $\beta$, $\gamma$, and $\delta$ lines have P Cygni profiles, which provides clear evidence of mass outflow. Close inspection of two He I lines at 1.083 $\mu$m and 2.058 $\mu$m also reveals P Cygni profiles, with weak emission components. From the blue edges of the P Cygni  profiles of the Paschen and He I lines we estimate a maximum outflow velocity of $\sim -550$ km/s; the absorption is centered at $\sim -400$ km/s. A weak emission line at 2.12 $\mu$m due to ${\rm H}_2$ may also be present, which would provide some evidence for shocked gas near MNO. However, we see no other signatures of ${\rm H}_2$ emission in our SpeX data.

The estimated magnitudes in the 2MASS photometric system at an airmass of one, as synthesized from our spectrum, are $J=10.90$, $H=8.85$ and $K_s=7.30$, with estimated uncertainties of about 10\%. The estimated magnitudes in the Mauna Kea Observatory photometric system (Simons \& Tokunaga 2002; Tokunaga, Simons, \& Vacca 2002) at an airmass of one are $J=10.83$, $H=9.00$, $K=7.20$, $L'=4.98$ and $M'=4.10$. These values are in reasonable agreement with the magnitudes estimated by Reipurth \& Aspin (2004) and suggest that the absolute flux calibration of our SpeX data is reliable. Comparison with the pre-outburst 2MASS magnitudes indicates that the source brightened by 3.84, 3.31, and 2.97 mags in $J$, $H$, and $K_s$ respectively, with an associated color change of $\Delta(J-H) = -0.53$ and $\Delta(H-K)=-0.34$ mags. Hence, in the NIR, MNO was bluer during outburst than in quiescence. 

The depth of the $3\mu$m ice feature implies an optical depth $\tau(3.0) \sim 0.7$. From the correlation between $\tau(3.0)$ and $A_V$ given by Whittet et al.\ (1988), we infer a reddening of $A_V \sim 11$. 
We have used this value and the reddening law of Rieke \& Lebofsky (1985) to correct the observed emission line fluxes for reddening. The observed fluxes, equivalent widths (EWs) and dereddened luminosities of many of the strongest features seen in our spectrum of MNO are given in Table 1.

The integrated flux in the 0.8$-$5.2 $\mu$m spectral range corresponds to a luminosity of $\sim 9.2 L_\odot$ before any reddening corrections. This is a factor of $\sim 3$ larger than the quiescent bolometric luminosity estimated by Lis et al\ (1999) from the combination of IRAS and millimeter data, and a factor of $\sim 2$ larger than that estimated by \'Abrah\'am et al.\ (2004) and Andrews et al.\ (2004) from integrating under the entire pre-outburst SED. Applying an extinction correction of $A_V = 11$, we find an intrinsic luminosity over this wavelength range of $\sim 27 L_\odot$.

Assuming that a $\sim 50 L_\odot$ increase in the bolometric luminosity during the outburst (Andrews et al.\ 2004) is due to accretion onto a $0.5 M_\odot$ pre-main sequence star and adopting the relation between radius and mass given by Kenyon, Calvet, \& Hartmann (1993), we estimate a mass accretion rate for MNO of $\sim 10^{-5} M_\odot$yr$^{-1}$ during the outburst. Using the outflow velocity derived from the P Cygni profiles, and assuming that HH23A represents a termination point for the outflow, we calculate a dynamical timescale for MNO of $\sim 800$ yr, significantly smaller than the value estimated by Eisl\"offel \& Mundt (1997) using tangential velocities derived from proper motions.

\section{Discussion}

Reipurth \& Aspin (2004) and Brice\~no et al.\ (2004) have suggested that MNO is a FUor or EXor object. However, our NIR spectrum of MNO does not resemble any existing NIR spectra of either type of object. Although a $K-$band spectrum of EX Lupi during outburst exhibited weak Br $\alpha$ in emission, the Na I, Ca I and the CO ro-vibrational $\Delta\nu=+2$ transitions were in absorption (Herbig et al.\ 2001).  In addition, the Balmer lines exhibited {\em reverse} P Cygni profiles (Lehmann, Reipurth, \& Brandner 1995). Similarly, the $^{12}$CO  transitions were in absorption in the $K-$band spectra of four known FUors (Reipurth \& Aspin 1997). While the brightness increase in MNO is larger than that for an EXor, our estimated accretion rate for MNO is well below the typical values found for FUors (Hartmann \& Kenyon 1996). Nevertheless, there are clear indications that MNO is a young stellar object. 

When the differences in reddening are taken into account, our 1.9$-$4.1$\mu$m spectrum of MNO is seen to be nearly identical to that of the Class I protostar SSV59 presented by Simon et al.\ (2004), although the emission features are somewhat weaker in the spectrum of SSV59. This supports the suggestion, based on the low pre-outburst bolometric luminosity (2.7$-$4.2 $L_\odot$; Lis et al.\ 1999; \'Abrah\'am et al.\ 2004; Andrews et al.\ 2004) and the submillimeter continuum slope (Andrews et al.\ 2004), that MNO is a late-stage Class I object. The fact that SSV59 also illuminates a reflection nebula and has a bolometric luminosity similar to that of MNO strengthens this conclusion.

The origin of H lines in the spectra of Class I sources is controversial. In models developed by Natta et al.\ (1988) and Hartmann \& Kenyon (1990) the H lines are generated in a magnetically driven, outflowing wind. However, based on the symmetric, purely emission profiles of Br$\gamma$ in the spectra of such sources, Najita et al.\ (1996) argued that the H lines are produced by mass infall, rather than outflow.  Furthermore, Muzerolle, Hartmann, \& Calvet (1998) used the Br$\gamma$ luminosity from Class I sources as a measure of the accretion luminosity and hence the mass accretion rate in these objects. However, neither the current magnetically driven wind models nor the magnetospheric accretion models are able to reproduce all of the various features of the H line profiles seen in the spectra of Class I sources (see Nisini et al.\ 2004 for a discussion). Nisini et al.\ (2004) have recently argued that the Br line strengths in the NIR spectrum of HH 100 result from an accelerating, ionized wind. 

The Paschen and two He I lines seen in our spectrum of MNO exhibit unmistakable P Cygni profiles. The strong H$\alpha$ line reported by Reipurth \& Aspin (2004) and Brice\~no et al.\ (2004) also exhibits a P Cygni profile. These profiles are clear indications of mass outflow in a wind. The Brackett emission lines, however, lack any absorption component and the centroid of the Br$\alpha$ emission is slightly blueshifted, consistent with the findings of Najita et al.\ (1996). This variation in H line profiles with recombination level is often seen in the spectra of hot massive stars with strong ionized and accelerating stellar winds, such as the B-type supergiants HDE 316285 (Hillier et al.\ 1998) and P Cygni itself (Najarro, Hillier, \& Stahl 1997; Lenorzer et al.\ 2002).  HDE 316285 also exhibits permitted lines of  Ca II and Na I in emission in its NIR spectrum. The estimated outflow velocity from MNO also matches the velocities seen in these systems. The similarities between the NIR spectra of these B supergiants and that of MNO provide support for the ionized wind model suggested by Nisini et al.\ (2004). Additional support is provided by the B-type optical spectrum of region `C' in the reflection nebula presented by Brice\~no et al.\ (2004). Presumably, the optical spectrum arises from the atmosphere of the disk itself, which may be driving the ionized wind.

Attempts to use the Pa and Br emission lines to corroborate the extinction estimate derived from the ice feature reveal a remarkable discrepancy with the standard Case B conditions for H I recombination (Hummer \& Storey 1987). The estimated $A_V$ value depends strongly on which line ratios are used, ranging from a minimum of 0.5 for the ${\rm Br} \gamma/{\rm Br}\alpha$ ratio to a maximum of 30 for the ${\rm Pa}\delta/{\rm Pa}\gamma$ ratio. The Br line ratios in particular suggest that the optically thin Case B assumption is violated for at least some of the low-lying H levels and transitions. A similar result has been reported for HH 100 by Nisini et al.\ (2004). We expect that some of the Balmer lines may also be optically thick, although clearly a full wind model is needed to determine the H level populations, from which the optical depths and line profiles can be predicted. High optical depth in the ionized wind would also explain why the 3.6, 6, and 20 cm radio emission has not increased during the outburst (Andrews et al.\ 2004). 

Under the assumption that the Br transitions are optically thick, their strengths provide an estimate of the size of the emitting regions. Following Nisini et al.\ (2004), we estimate the size of the Br emission region by assuming the line flux is given by blackbody emission and adopting a gas temperature of $10^4$ K. We find that the Br $\alpha$ emission region has a radius $\sim 8.5 R_\odot$. From the correlations between mass loss rate and bolometric luminosity, and between mass loss rate and Br $\alpha$ luminosity,  presented by Nisini et al.\ (1995), the inferred mass loss rate for MNO is of the order of $4 \times 10^{-8} M_\odot$yr$^{-1}$, substantially below the typical values found for FUor objects (Hartmann \& Kenyon 1996). From these values and the wind velocity determined from our spectrum, we estimate a density in the emission region of the order of $10^{9-10}$ cm$^{-3}$. For such a dense wind, collisions will play a significant role in determining the populations of the low-lying H levels and excited-state photoionization will be important in maintaining the ionization equilibrium (see e.g., Simon et al.\ 1983). A similar explanation has been suggested for the H emission lines in the spectrum of the embedded B1.5Ve star MWC 297, which may also possess a disk that drives an outflow (Drew et al.\ 1997). At these high densities, the forbidden lines normally seen in photoionized regions may be collisionally de-excited, which would explain why they are not seen in the optical spectrum of MNO (Reipurth \& Aspin 2004; Brice\~no et al.\ 2004).

Although the connection between MNO and other eruptive objects (FUors and/or EXors) is unclear at the moment, the conclusion that MNO is a Class I object seems secure. It is possible that the picture we have outlined for MNO may be applicable to other Class I sources in L1630, and MNO may therefore represent a natural evolutionary transition from theÊ deeply embedded Class 0 objects in nearby
areas of the same cloud to the visible Class II T Tauri stars that are also present in the vicinity.

\acknowledgments
WDV thanks G\"oran Sandell for many useful discussions and Sean Andrews and Barry Rothberg for sharing their results prior to publication. We thank Dave Griep and Bill Golisch for their assistance at the telescope. This work was supported by the National Aeronautics and Space Administration under Cooperative Agreement no. NCC 5-538 issued through the Office of Space Science, Planetary Astronomy Program.


\clearpage

\begin{deluxetable}{llcrc}
\tablecaption{Line fluxes, EWs, and luminosities  in the NIR spectrum of MNO from 2004 March 09, 06h UT.}
\tablecolumns{5}
\tablewidth{0pc}
\tablehead{
\colhead{Line} & \colhead{Wavelength}& \colhead{Flux\tablenotemark{a}} & \colhead{EW\tablenotemark{a}} & \colhead{Luminosity\tablenotemark{b}}\\
\colhead{}    & \colhead{($\mu$m)}  & \colhead{(${\rm W}/{\rm m}^2$)} & \colhead{(\AA)} & \colhead{(W)} }
\startdata
Ca II & 0.8498 & $1.6 \times 10^{-17}$  & $-9.8$ &$6.7 \times 10^{24}$ \\
Ca II & 0.8542 & $2.1 \times 10^{-17}$ & $-12.4$ & $8.2 \times 10^{24}$\\
Ca II & 0.8662  & $1.8 \times 10^{-17}$ & $-10.1$ & $6.1 \times 10^{24}$\\
Pa $\delta$ \tablenotemark{c} &  1.0049    & $2.7 \times 10^{-18}$  & $-0.6$ & $2.5 \times 10^{23}$\\
He I \tablenotemark{d} & 1.0830   & \nodata                      & $8.9$ & \nodata \\
Pa $\gamma$\tablenotemark{c} & 1.0938 & $2.1 \times 10^{-17}$  & $-2.8$ & $1.2 \times 10^{24}$\\
O I (?) &  1.1287   & $1.6 \times 10^{-17}$ & $-1.8$ & $7.8 \times 10^{23}$ \\
Pa $\beta$ \tablenotemark{c}       & 1.2818 & $1.7 \times 10^{-16}$ & $-10.9$ & $5.0\times 10^{24}$\\
Mg I & 1.5031     & $5.4 \times 10^{-17}$ & $-2.2$ &$7.9 \times 10^{23}$ \\
Pa $\alpha$ \tablenotemark{c} &  1.8751  & $2.1 \times 10^{-15}$ & $-52$ & $1.5 \times 10^{25}$\\
Br $\delta$  + Ca I  & 1.9446 & $1.4 \times 10^{-16}$ & $-3.4$ & $9.6 \times 10^{23}$\\
He I \tablenotemark{d} & 2.0581  &  \nodata                      & $0.7$ & \nodata \\
Br $\gamma$  & 2.1655 & $2.3 \times 10^{-16}$ & $-4.6$ & $1.4 \times 10^{24}$\\
Na I & 2.2084 & $4.8 \times 10^{-17}$ & $-0.9$ &$2.9 \times 10^{23}$ \\
Br $\alpha$ &  4.0512   & $6.6 \times 10^{-16}$ & $-12$ & $1.9 \times 10^{24}$\\
CO & 4.67 & \nodata & $65$ & \nodata\\
\enddata
\tablenotetext{a}{Estimated errors are 10\% in both flux and EW.}
\tablenotetext{b}{Dereddened luminosity for $A_V = 11$ and $d=400$ pc.}
\tablenotetext{c}{P Cygni profile. Flux and EW refer to emission component only.} 
\tablenotetext{d}{P Cygni profile. Flux and EW refer to absorption component only.} 
\end{deluxetable}

\clearpage
\begin{figure}
\hspace{-0.9in}
\begin{minipage}[c]{4.5in}
\centerline{\hbox{\psfig{file=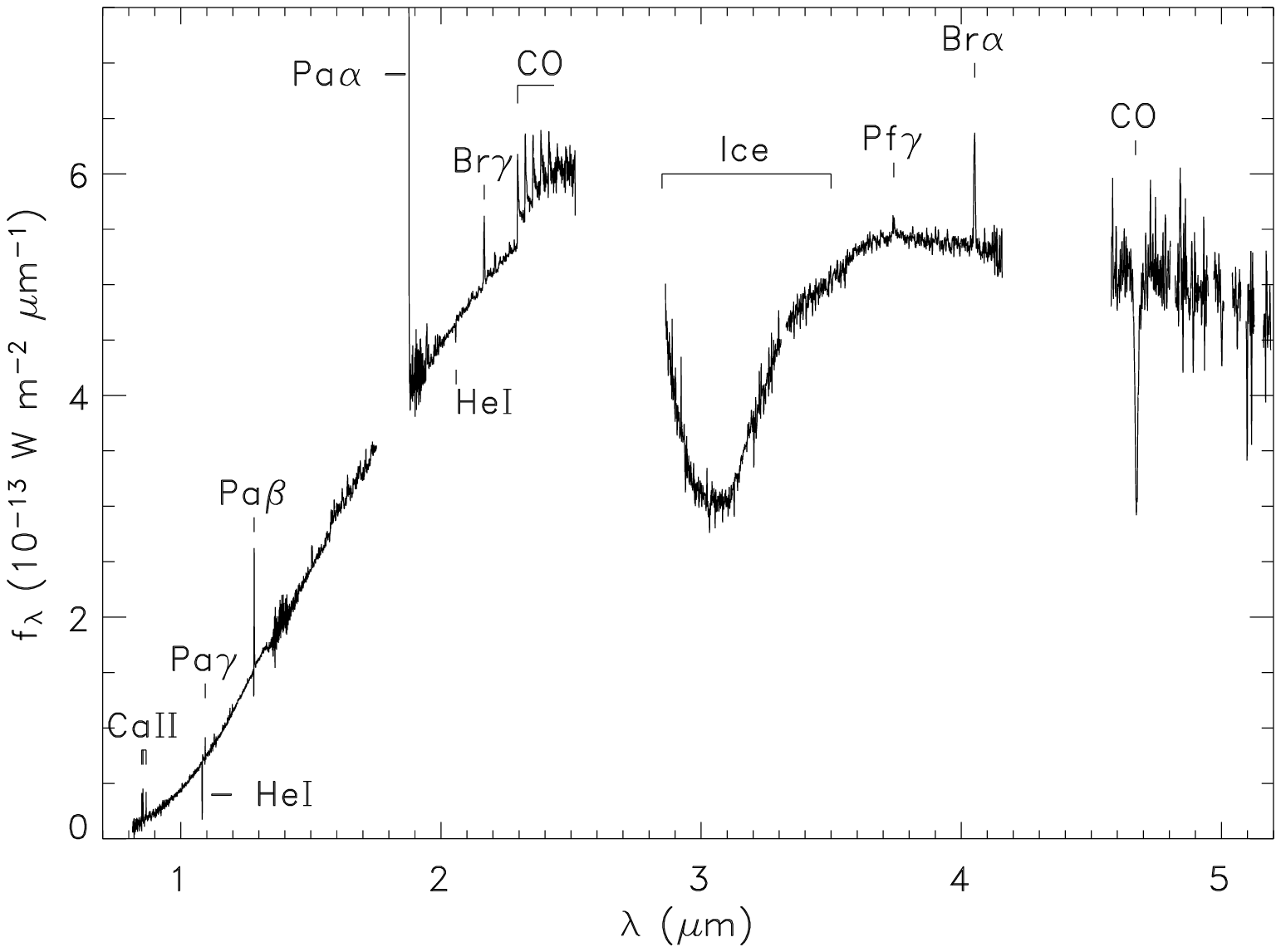,width=4.5in,angle=0}}}
\end{minipage}
\begin{minipage}[c]{3.5in}
\centerline{\hbox{\psfig{file=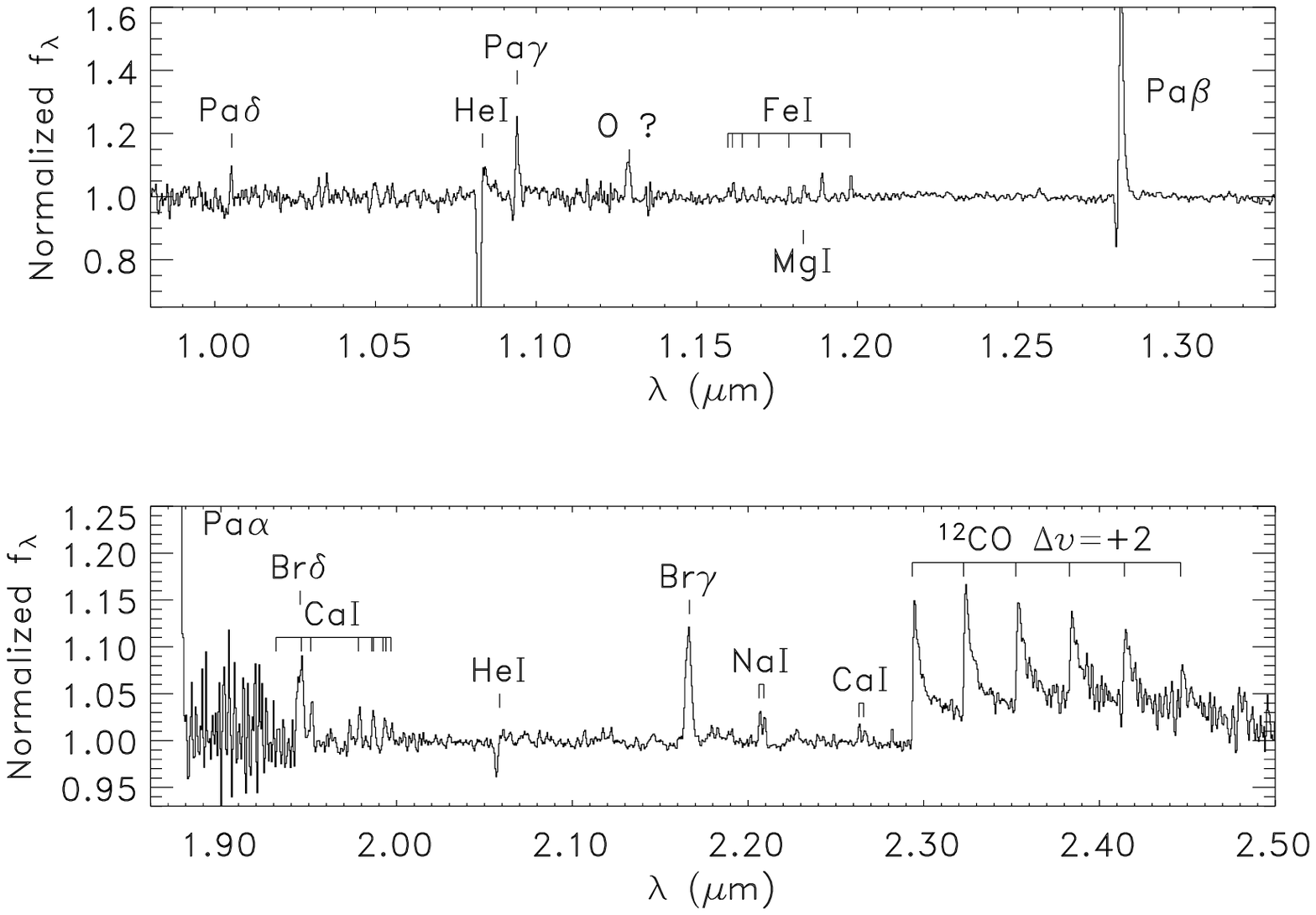,width=3.75in,angle=0}}}
\end{minipage}
\caption{({\em left})The 0.8$-$5.2$\mu$m spectrum of McNeil's Nebula Object obtained with SpeX. Prominent emission and absorption features are labeled. ({\em right}) Normalized J- and K-band spectra of MNO.} 
\end{figure}

\end{document}